\pdfoutput=1
\PassOptionsToPackage{unicode}{hyperref}
\documentclass[11pt,final]{article}

\usepackage{authblk}
\usepackage{amsfonts}
\usepackage{amsmath}
\usepackage{listings}
\usepackage[svgnames]{xcolor}
\usepackage{yfonts}
\usepackage{graphicx}
\usepackage{verbatim}
\usepackage{bm}
\usepackage[]{algorithm2e}
\usepackage{caption}
\usepackage{import}
\usepackage{tikz}

\usepackage[utf8]{inputenc}
\usepackage[T1]{fontenc}
\usepackage{lmodern}
\usepackage{textgreek}

\usepackage{varioref}
\usepackage{xspace}
\usepackage{enumitem}

\lstdefinelanguage
[rv64gc]{Assembler}
{alsoletter={.},
	morekeywords=
	{sra,srli,mv,xor,sw,
		addi,addiw,c.li,c.lui,fld,ld,li,lui,lw,amoand.d,amoand.d.aq,amoor.w.aq,
		amoand.d.aqrl,amoand.d.rl,amoand.w,amoand.w.aq,amoand.w.aqrl,
		amoand.w.rl,amoor.d,amoor.d.aq,amoor.d.aqrl,amoor.d.rl,amoor.w,
		amoor.w.aq,amoor.w.aqrl,amoor.w.rl,bgtz,ble,bleu,blez,blt,bltu,
		csrc,csrci,csrr,csrrc,csrrci,csrrs,csrrsi,csrrwi,csrwi,fcvt.d.q,
		fmadd.d,fmadd.q,fmadd.s,fmsub.d,fmsub.q,fmsub.s,fnmadd.d,fnmadd.q,
		fnmadd.s,fnmsub.d,fnmsub.q,fnmsub.s,j,jal,lui,sra},
	morecomment=[l]{\#},
	morecomment=[s]{/*}{*/}}

\newcommand*{\alpvar}[2]{\texttt{{#1}RV64{#2}}\xspace}
\newcommand*{\alpslash}{\alpvar{/}{IAC}}
\newcommand*{\alphash}{\alpvar{\#}{IC}}
\newcommand*{\alptick}{\alpvar{'}{IDC}}
\definecolor{mygray}{RGB}{210,210,210}

\usepackage[backend=bibtex,style=numeric]{biblatex}
\PassOptionsToPackage{hyphens}{url}
\usepackage{hyperref}
\usepackage{cleveref}

\begin{document}
\date{}
\author[1]{Hadrien Barral}
\author[1]{R\'emi G\'eraud-Stewart}
\author[1,2]{Georges-Axel Jaloyan}
\author[1]{David Naccache}
\affil[1]{DIENS, \'Ecole normale sup\'erieure, CNRS, PSL University, Paris, France}
\affil[2]{CEA, DAM, DIF, F-91297 Arpajon, France}

\title{RISC-V: \texttt{\#}AlphanumericShellcoding}

\maketitle
\thanks{Modified version of our paper that appeared at WOOT 2019, co-located with USENIX Security.}
	\begin{abstract}
	We explain how to design RISC-V shellcodes capable of running arbitrary code, whose ASCII binary representation use only letters \texttt{a--zA--Z}, digits \texttt{0--9}, and either of the three characters: \texttt{\#}, \texttt{/}, \texttt{'}.
\end{abstract}

\section{Introduction}

RISC-V~\cite{RISCV} is a new \emph{Instruction Set Architecture} (ISA) which development began in 2010. It is based on the concept of \emph{Reduced Instruction Set Computer} (RISC)~\cite{RISC1}, targeting simplicity by providing few and limited computer instructions. RISC ISAs have become increasingly popular with the wide adoption of embedded devices such as smartphones, tablets, or other Internet of Things devices. The most popular RISC ISAs are currently ARM~\cite{ARMv8A}, Atmel AVR~\cite{AtmelAVR}, MIPS~\cite{MIPS}, Power~\cite{Power}, and SPARC~\cite{SPARC}.

RISC-V is the fifth RISC ISA published by UC Berkeley. It is completely free and open-source, with its User-Level ISA published in May 2017 in version 2.2. It features 32-bit and 64-bit little-endian variants (designated as \texttt{RV32} and \texttt{RV64}), with a future extension to 128-bit. While only test boards feature RISC-V processors for now, many companies including Western Digital or Nvidia have announced the use of RISC-V chips in their future products~\cite{westerndigital}.

This increasing popularity makes RISC-V an attractive target for low-level attackers, whose tools are traditionally honed chiefly against x86 platforms. This is exacerbated by the widespread adoption of smartphones and their use for almost any task from payment to dating to unlocking one's car, making successful attacks especially profitable. At the same time, mobile environments are improving their security, in part to address such threats.

Nevertheless, mobile applications are still occasionally vulnerable to memory safety vulnerabilities, and the need for performance or obfuscation often drives developers to implement low-level (e.g., JNI) segments which are particularly susceptible to the usual techniques of buffer overflow exploitation. Unlike traditional local or network scenarios however, the attacker has only limited ways to transmit a payload.

We claim that a reasonable vector is text-based applications, which includes SMS, social networks, chat applications (in a remote context), password entry, note taking, or QR code scanning (in a local context). This being said, the attacker's payload has to be treated by this application as text such as a hashtag, a URL, a sentence, in the most restrictive sense, hence the most widely applicable. We therefore consider alphanumeric programs whose binary representation use only the alphanumeric ASCII characters: the 52 lowercase and uppercase letters of the English alphabet and the 10 digits. As we will discuss, it is only possible to achieve \emph{arbitrary} code execution at the cost of allowing one additional character: either \texttt{\#}, \texttt{/}, or \texttt{'}, each being compatible with the use-cases discussed above.

\subsection{Prior and related work}

This work follows a trend initiated in the early 2000s to evade buffer overflow protections
(Eller~\cite{eller2000} and RIX~\cite{RIX2001} on x86) and intrusion detection systems~\cite{DBLP:conf/ccs/MasonSMM09}.
Tools to generate alphanumeric shellcodes on the x86 platform appeared \cite{DBLP:conf/iciss/BasuMC14} and are now a standard component of attack frameworks including Metasploit (\texttt{msfvenom}) and UPX1. The x86 is particularly well suited to this exercise as many letters materialize into \texttt{mov} instructions, which form a Turing-complete subset of operations \cite{MovTC}.
To this day however neither of these tools are able to generate alphanumeric shellcodes on ARM platforms, such as ARMv8 and RISC-V.

The first automated tool was provided by Younan et al. in 2011 for the ARMv5 platform, relying on an BF interpreter and bytecode \cite{Younan}. The technique however does not carry over to more recent implementations. In 2016,
Barral et al. introduced the first tool capable of compiling arbitrary ARMv8
code into alphanumeric executable code \cite{AlphanumericARMv8}. This is a \emph{tour de force} but also and most importantly it introduces a generic approach to design such tools.

To the best of the authors' knowledge, none of the currently available approaches works on RISC-V.

\subsection{Our contribution}

We provide, as far as the authors know, the first analysis of alphanumeric code on RISC-V, as well as a complete framework for automatically generating alphanumeric (+1 character) shellcodes. Through a three-staged modular design, these shellcodes achieve arbitrary code execution on this platform.

This is the second architecture which can be addressed using the methodology from \cite{AlphanumericARMv8}, which is an argument in favor of such generic approaches (rather than \textit{ad hoc} ones). Our approach differs on the fact that we do not manually assemble available instructions into higher-level constructs for building the unpacker in a bottom-up fashion and instead opt for a partially automated strategy to generate the required alphanumeric instruction sequences to achieve the desired results.

We provide three different constructions, corresponding to each choice of an additional character. All our programs are given in appendix, being to the best of the authors' knowledge the first automated tool of this kind for RISC-V, as well as the first examples of such shellcodes for each construction.

\section{Background}
\subsection{Shellcodes and exploitation}
In a typical arbitrary code execution (ACE) scenario, attackers can run a relatively short program of their choosing.
It is called a \emph{shellcode}, as it can start a shell session, which in turn allows attackers to download and run
additional programs.

For instance, a stack overflow ACE can happen when an application allows writing in an array beyond the allocated space
for this array, resulting in overwriting stack frame data. In platforms such as x86 the stack frame stores information
about the instruction pointer before a call; by overwriting this information an attacker can control the instruction pointer
and send it back to the array's address. The array's contents are then executed as if they were the vulnerable program's own
instructions: this is where the shellcode is written.

Since a typical array is relatively short, shellcodes must accordingly be concise.
Similarly, an application may restrict what data it manipulates (e.g., strings) and shellcodes must be written to comply with such constraints.
Additional protections make shellcode design trickier: \emph{Address Space Layout Randomization} (ASLR), stack-smashing protections, or non-executable stack space for instance.
Detection mechanisms may furthermore identify characteristic aspects of a shellcode and prevent the attack from reaching the target application.
For all these reasons the modern shellcode designer has to navigate around layers of obstacles.

This difficulty is somewhat offset on embedded and mobile devices, where many protections are only partially implemented, if at all.
Such platforms are also host to many third-party applications, that can be developed without strictly adhering to secure coding practices,
using memory-unsafe languages (sometimes due to performance or obfuscation constraints) and not necessarily updated in a timely fashion.

\subsection{The RISC-V instruction set}

RISC-V splits its instruction set between a mandatory core set (\texttt{RV64I}) and different optional extensions, each of which is designated by a string (a single letter for the most common ones), among which integer multiplication and division~(\texttt{M}), atomic operations~(\texttt{A}), single-, double- or quad-precision~(\texttt{F}, \texttt{D}, \texttt{Q}) floating-point operations, decimal floating-point operations~(\texttt{L}), compressed instructions~(\texttt{C}), vector operations~(\texttt{V}).

The general purpose ISA, which includes \texttt{IMAFD}, is designated by the letter \texttt{G}. In what follows, we focus on the \texttt{RV64GC} ISA, which is the one agreed on by Debian and Fedora porters, as well as members of the RISC-V Foundation. On top of that, the Foundation intends to provide ``\textit{a profile for standard RISC-V Unix platforms that will include \texttt{C} extension as mandatory}''~\cite{debian}.

The \texttt{RV64GC} ISA features 32-bit and 16-bit instructions, aligned on 16 bits. There are 31 general purpose 64-bit registers (\texttt{x1}-\texttt{x31}), 32 floating-point registers (\texttt{f0}-\texttt{f31}), a program counter (\texttt{pc}), as well as various control-and-status registers. The pseudo-register \texttt{x0} designates the zero constant.

We adopt for the rest of this paper some terminology defined by the RISC-V Instruction Set Manual, Version 1.10~\cite{RISCV}. Assembly instructions are written in the format \texttt{add x1,x2,x3}, where \texttt{add} is called the \emph{opcode}, and \texttt{x1}, \texttt{x2}, \texttt{x3} are the \emph{operands}. Precisely, \texttt{x1} is the \emph{destination register}, \texttt{x2} is the \emph{first source register} and \texttt{x3} is the \emph{second source register}. When one of the source registers is replaced by a constant, it is called an \emph{immediate}. To those conventions, let $K$ be a register, we add our slicing notation as $K[y:x]$ (with $x < y$), meaning we take a slice of bits $x$ to $y$ of $K$, with the lowest bit denoted as the bit 0.

RISC-V ELF psABI specification~\cite{RVELFABI} provides a register naming convention, reproduced in \Cref{Fig:psABI}.

\begin{figure}[h]
	\centering
	\begin{tabular}{|l|c|l|}
		\hline
		Register & ABI Mnemonic & Meaning \\
		\hline
		\texttt{x0} & \texttt{zero} & Zero \\
		\texttt{x1} & \texttt{ra} & Return address \\
		\texttt{x2} & \texttt{sp} & Stack pointer \\
		\texttt{x3} & \texttt{gp} & Global pointer \\
		\texttt{x4} & \texttt{tp} & Thread pointer \\
		\texttt{x5}-\texttt{x7} & \texttt{t0}-\texttt{t2} & Temporary registers \\
		\texttt{x8}-\texttt{x9} & \texttt{s0}-\texttt{s1} & Callee-saved registers \\
		\texttt{x10}-\texttt{x17} & \texttt{a0}-\texttt{a7} & Argument registers \\
		\texttt{x18}-\texttt{x27} & \texttt{s2}-\texttt{s11} & Callee-saved registers \\
		\texttt{x28}-\texttt{x31} & \texttt{t3-}\texttt{t6} & Temporary registers \\
			\hline
			\hline
			\texttt{f0}-\texttt{f7} & \texttt{ft0}-\texttt{ft7} & Temporary registers \\
			\texttt{f8}-\texttt{f9} & \texttt{fs0}-\texttt{fs1} & Callee-saved registers \\
			\texttt{f10}-\texttt{f17} & \texttt{fa0}-\texttt{fa7} & Argument registers \\
			\texttt{f18}-\texttt{f27} & \texttt{fs2}-\texttt{fs11} & Callee-saved registers \\
			\texttt{f28}-\texttt{f31} & \texttt{ft8}-\texttt{ft11} & Temporary registers \\
		\hline
		
	\end{tabular}
	\captionof{table}{Naming convention for registers, per psABI~\cite{RVELFABI}.}
	\label{Fig:psABI}
\end{figure}

\section{Alphanumeric RISC-V}
The first step towards building an alphanumeric shellcode for \texttt{RV64GC} consists in generating the subset of alphanumeric valid instructions, which we denote by $\alpha\texttt{RV64GC}$. For this purpose, we generated every 16-bit and 32-bit alphanumeric sequence, and tentatively disassembled it using \texttt{objdump}. Per RISC-V Instruction Set Manual, 16-bit instructions must have their two least significant bits set to \texttt{00}, \texttt{01} or \texttt{10}. Similarly, 32-bit instructions must have their five least significant bits set to \texttt{bbb11}, with \texttt{bbb} different from \texttt{111}.

Furthermore, some opcodes may encode invalid or unimplemented instructions. For instance, the little-endian word \texttt{7OOT} corresponds to a load upper immediate (\texttt{lui}), whereas \texttt{WOOT} does not correspond to any valid \texttt{RV64GC} instruction, although its least significant bits are those of a valid 32-bit instruction:

\begin{lstlisting}
7OOT   0x374f4f54   lui t5,0x544f4
WOOT   0x574f4f54   undefined
\end{lstlisting}

After filtering out all invalid sequences, we regroup the remaining instructions according to their opcode, providing an overview of the available instructions for which there are some operands making them alphanumeric.

The internal structure of the instruction defines the main constraints on the alphanumeric language subset. Each 32-bit instruction has its opcode encoded in the first 7 bits of the first byte. Requiring the first byte to be alphanumeric will therefore greatly reduce the available opcodes, while providing a wide range of operands for each opcode. On the contrary, 16-bit instructions are more entropic in their spread. Henceforth, more opcodes are available, with less operands for each opcode. Consequently, the expressiveness of $\alpha\texttt{RV64GC}$ relies on the intelligent combination of instructions of various lengths.

Hereafter, we provide a review of those instructions, by explaining their semantics and some insight on the available operands. For simplicity and following the methodology introduced by Barral et al. in \cite{AlphanumericARMv8}, we cluster instructions as \emph{control-flow}, \emph{data processing}, and \emph{memory manipulation} instructions.

\subsection{Data processing}
Data processing includes every instruction that does not modify the memory or the program counter. Two variants may be available for each instruction, either operating on the usual 64-bit registers or performing the operation on 32 bits and sign-extending the result to the 64-bit register. Using 32-bit variants for pointer manipulation prevents from reaching addresses ranging from \texttt{0x8000~0000} to \texttt{0xFFFF~FFFE~FFFF~FFFF}. This is a serious caveat for bare-metal shellcodes --- as existing boards often have the DRAM start at \texttt{0x8000~0000} --- forcing us to use the 64-bit variant. Hereafter, we only present the most useful ones, omitting instructions which may have odd effects (like micro-architectural hints for branch predictors):
\begin{itemize}	
	\item The addition \texttt{addi} instruction enables adding or removing only some specific values multiple of 16 to \texttt{sp}. Its 32-bit signed variant \texttt{addiw} is also available, and allows increasing or decreasing registers \texttt{a0}, \texttt{a2}, \texttt{a4}, \texttt{a6}, \texttt{s0}, \texttt{s2}, \texttt{sp}, \texttt{t1}, and \texttt{tp} by a value ranging from $-20$ to $30$.
	
	\item The instruction \texttt{li}, allows loading signed integers (ranging from $-20$ to $30$) into registers \texttt{sp}, \texttt{tp}, \texttt{s0}, \texttt{s2}, \texttt{s4}, \texttt{s6}, \texttt{s8}, \texttt{s10}, \texttt{t1}, \texttt{t3}, \texttt{t5}, \texttt{a0}, \texttt{a2}, \texttt{a4}, and \texttt{a6}. We use the letter $\mathcal{S}$ to designate this set of registers.
	
	Loading immediates to registers may also be done with the \texttt{lui} instruction (load upper immediate), which loads a 20-bit signed immediate into the bits 31-12 of a register in $\mathcal{S}$. The lowest bits are all set to zero, while the 32 highest bits are computed as the sign-extension of the immediate. We counted 238,791 alphanumeric \texttt{lui} instructions, with a large choice of immediates.
	
	\item Bitwise manipulation: only the \texttt{sra} (shift right arithmetical) instruction is available, with all registers of $\mathcal{S}$ as source and destination, and registers \texttt{s3}--\texttt{s7} as shift amount.
	
	\item Floating-point operations: many useful floating-point operations are available in $\alpha\texttt{RV64GC}$, in simple, double and quad precision. Among them we find sign manipulation like \texttt{fabs} (absolute value) or multiply-accumulate \texttt{fmadd} and its variants ($r \leftarrow \pm a \times b \pm c$).
	
	\item Control-status register manipulation: many instructions available, such as \texttt{csrc}, \texttt{csrci}, \texttt{csrrc}, \texttt{csrrci}, \texttt{csrrsi}, \texttt{csrrwi}, \texttt{csrwi}, not detailed here. As privileged access may be required, we preferred not using them and instead use the other available data processing instructions.
	
\end{itemize}

\subsection{Control-flow instruction}
Both conditional and unconditional jump instructions are available. For unconditional branching, we have both \texttt{j} (jump) and \texttt{jal} (jump and link) available, with the possibility to link any register of $\mathcal{S}$. Conditional branches are also available, with a wide variety of branching conditions: \texttt{bgtz}, \texttt{ble}, \texttt{bleu}, \texttt{blez}, \texttt{blt}, \texttt{bltu}. No backward jump is available, as the immediate offset has its sign set on the highest bit of a byte (hence always equal to zero when alphanumeric). This may prevent the Turing completeness of $\alpha\texttt{RV64GC}$, as no unbounded computation mechanism is available without additional assumptions, such as code-reuse or self-modifying code.

\subsection{Memory processing}
We have both 32-bit \texttt{lw} and 64-bit \texttt{ld} loads as well as double-precision floating-point \texttt{fld} loads. However, no stores are available, which makes it impossible to write arbitrary shellcodes: we are only able to modify the registers and not the machine's memory state.

This turns out to be a strong limitation as for instance the shellcode designer cannot build paths (such as \texttt{"/bin/sh"}) in memory (this is not an alphanumeric string). Thus, additional assumptions must be made, either by finding gadgets able to write to memory or by reusing memory previously set to the desired value at a known position (e.g., an environment variable). Either option seems unsuitable in the context of a self-contained shellcode.

We therefore consider the possibility of allowing one non-alphanumeric character --- a choice which may be governed by operational constraints as well.
Among all ASCII-printable instructions modifying memory, only three non alphanumeric characters stand out: slash~\texttt{/}, hash~\texttt{\#}, and tick~\texttt{'}.

\begin{itemize}	
	\item Adding the hash character \texttt{\#} gives standard 32-bit \texttt{sw} and 64-bit \texttt{sd} store instructions. The 32-bit store \texttt{sw} provides the ability to store almost any variable to various addresses with offsets multiple of 32. Given that there is no possibility to increment a 64-bit register by less than 16 (using \texttt{addi}), many memory areas are out of reach. The 64-bit variant \texttt{sd} seems more promising: indeed, the available offsets are only 2 bytes apart. Using this, we can efficiently store data by using \texttt{addi} increments for coarse-grained pointer manipulation and reaching the exact store address (up to a precision of 2 bytes) by tweaking the offset of \texttt{sd}.
	
	\item Adding the slash character \texttt{/} provides some atomic instructions, such as 32-bit and 64-bit atomic read-modify-write variants of binary conjunction \texttt{amoand} and disjunction \texttt{amoor}. As an example, \texttt{amoor.d t1,s5,(sp)} loads 64 bits from the address in \texttt{sp} into \texttt{t1}, and stores in the same address the disjunction of \texttt{t1} and \texttt{s5}. Note that the addresses passed to atomic operations must be naturally aligned, which adds further complexity when designing our shellcode.
	
	\item Adding the tick character \texttt{'} provides floating-point store instructions \texttt{fsd}, \texttt{fsq}, \texttt{fsw}. Controlling the stored values requires deep technical knowledge of floating-point binary representation, as the associated data manipulation operations are of the form $\pm a \times b\pm c$ (e.g., \texttt{fmadd}, \texttt{fmsub}).
\end{itemize}
For each of these three characters, we define a new subset of \texttt{RV64GC}, denoted respectively \alphash, \alpslash and \alptick. The following section details how we can achieve ACE in \alphash, setting up the stage and much of the machinery for shellcodes in \alpslash and \alptick as well. Since these require additional work, they are discussed later on.

\section{High-level design}
\label{sec:highlevel}
Several approaches can be used to run arbitrary code from an instruction-limited shellcode. The main techniques available are: virtualization, compilation, and packing.

Virtualization, as used by Younan et al. for 32-bit ARMv7 alphanumeric shellcoding~\cite{Younan}, requires the design of a bytecode and an interpreter, both compatible with the limited instruction set, and powerful enough to mount a realistic attack --- beyond Turing-completeness, we need to perform system calls or other mechanisms to evade the virtual environment. Virtualization presents a huge runtime overhead as well as a committed engineering effort.

Compilation, when applicable, is very efficient: compilers such as \emph{movfuscator}~\cite{MovTC,Movfuscator} and \emph{higher subleq}~\cite{Subleq} have been provided for \emph{one instruction set computers}, reduced ISA subsets made of only one instruction. However, such methods are not applicable to $\alpha\texttt{RV64GC}$ as they often rely on \emph{syntax-directed translation schemes}. Here, the heavy constraints in $\alpha\texttt{RV64GC}$ on the instruction operands hinders such methods that systematically translate each grammar symbol into the target language. Furthermore, writing compilers is in itself a daunting task. Perhaps for these reasons, to the best of our knowledge, no work on compilation for alphanumeric shellcoding has been published.

Packing is the third method, and by far the most common approach in shellcoding. This typically results in multi-staged shellcodes, where one stage decodes a second stage which is then executed. Packers can provide additional functionalities such as encryption, which we do not explore here. However, this technique requires the ability to execute self-modifying code, which may be hindered by the presence of \emph{executable space protection} mechanisms like DEP~\cite{DEP}, PaX~\cite{PaX} or NX-bit~\cite{NXbit}. Moreover, self-modifying code raises cache issues which need to be handled on a target-specific basis.

We decided to follow this third approach: it is conceptually simpler, much easier to check for correctness, and well suited to our target platform.

\begin{figure}[ht]
	\centering
	\begin{tikzpicture}
	
	\fill [mygray] (-1.5,-5) rectangle (1.5,-7);
	\node at (0, 0.5) {\textsc{Stage 1}};
	\draw (-1.5,0) rectangle (1.5, -7); 
	\draw (-1.5,-1) to (1.5, -1);
	\draw (-1.5,-2) to (1.5, -2);
	\draw (-1.5,-4) to (1.5, -4);
	\draw (-1.5,-5) to (1.5, -5);
	\draw (-1.5,-7) to (1.5, -7);

	\node at (0,-0.5) {init};
	\node at (0,-1.5) {forward jump};
	\node at (0, -2.75) {encoded payload};
	\node at (0, -3.25) {$\mathcal P_\text{enc}$};
	\node at (0, -4.5) {Unpacker $\mathcal U_1$};
	\node at (0, -5.75) {\textsc{Stage 2} $\mathcal U_2$};
	\node at (0, -6.25) {(unpacked by $\mathcal U_1$)};
	\draw[>=latex,->,rounded corners=3mm, very thick] (1.5,-1.5) -- (2,-1.5) -- (2, -4.5) -- (1.5,-4.5);
	\end{tikzpicture}
	\caption{General structure of stage 1: an initialization section, with a forward jump over the data-pool that contains the encoded final payload $\mathcal{P}_{enc}$, and the unpacker $\mathcal{U}_1$. The location at which the stage 2 is unpacked is highlighted in gray.}
	\label{fig:stage1}
\end{figure}
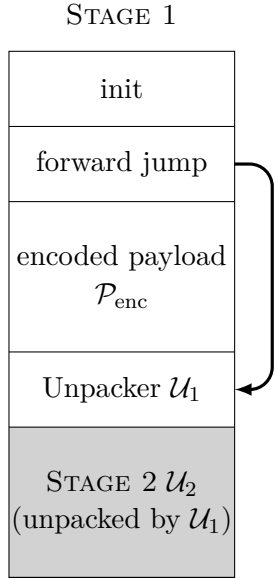
\begin{figure}[ht]
	\centering
	\begin{tikzpicture}
	\usetikzlibrary{decorations.pathreplacing}
	
	\fill [mygray] (-1.5,-6.5) rectangle (1.5,-8.5);
	\node at (0, 0.5) {\textsc{Stage 2}};
	\draw (-1.5,0) rectangle (1.5, -5); 
	\draw (-1.5,-1) to (1.5, -1);
	\draw (-1.5,-3) to (1.5, -3);
	\draw (-1.5,-4) to (1.5, -4);
	\draw (-1.5,-5) to (1.5, -5);
	\draw (-1.5,-6.5) rectangle (1.5, -8.5);
	\node at (0,-0.5) {init};
	\node at (0, -1.5) {main};
	\node at (0, -2) {decoding};
	\node at (0, -2.5) {loop};
	\node at (0, -3.5) {backward jump};
	\node at (0, -4.5) {jump};
	\node at (0, -7.25) {\textsc{Stage 3}};
	\node at (0,-7.75) {Payload};
	\draw[>=latex,->, very thick] (0,-5) to node[midway,fill=white]{\raisebox{0pt}[0.65\height][0.0\height]{$\vdots$}} (0,-6.5);
	\draw[>=latex,->,rounded corners=3mm, very thick] (1.5,-3.5) -- (2,-3.5) -- (2, -1) -- (1.5,-1);
	\draw [decorate,decoration={brace,amplitude=10pt,mirror,raise=4pt},yshift=0pt]
	(-1.5,0) -- (-1.5,-5) node [black,midway,xshift=-0.8cm] {$\mathcal U_2$};
	\end{tikzpicture}
	\caption{General structure of stage 2: an initialization section, with a loop decoding at each iteration one byte of the final payload $\mathcal{P}$ using two bytes of the encoded payload $\mathcal{P}_{enc}$. It finally jumps to the decoded payload, highlighted in gray.}
	\label{fig:stage2}
\end{figure}
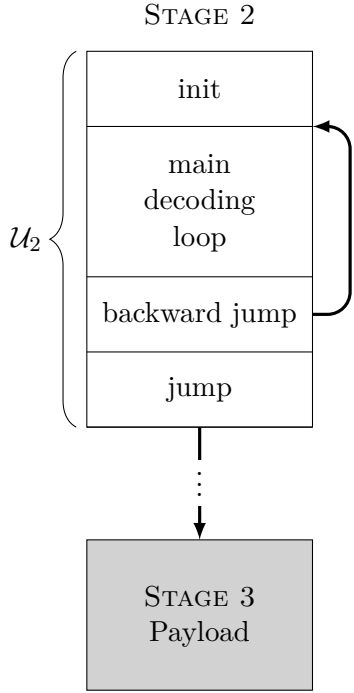

\section{Detailed construction}
\label{sec:Implementation}
In this section we show how to achieve arbitrary code execution, by detailing each step of the \alphash version of the shellcode. Building on the foundations laid with \alphash, we achieve similar results in \alpslash and \alptick.

\pagebreak

As explained in \Cref{sec:highlevel}, we use a packing multi-staged design. We present a three-stage approach:
\begin{itemize}	
	\item The first stage is a \emph{specific} unpacker written in \alphash;
	\item The second is a \emph{general} unpacker written in a slightly larger subset of \texttt{RV64IC};
	\item The third is our arbitrary payload.
\end{itemize}
The rationale for using three stages is governed by \alphash not containing backjumps, therefore forcing us to unroll the decoding logic. This would result in unwieldy large shellcodes if there were only two stages. Instead, we use the first unpacker $\mathcal{U}_1$, whose structure is shown in \Cref{fig:stage1}, to unpack a minimal program $\mathcal{U}_2$ shown in \Cref{fig:stage2}.
The program $\mathcal U_2$ \emph{has backward jumps} and can therefore efficiently implement a decoder using a loop. $\mathcal U_2$ unpacks and execute the third stage, which is the payload $\mathcal{P}$.

\subsection{Stage 1}
$\mathcal{U}_1$ is an unpacker for the next stage. It is fully written in \alphash. As no backward jumps are available, the unpacker is written as a \emph{straight-line program}.

Specifically, $\mathcal{U}_1$ must: (1) locate the shellcode and jump over the encoded payload; (2) fix-up the store pointer; (3) unpack stage 2; (4) jump to the decoded stage 2.

We achieve (4) simply by placing the decoded stage 2 immediately after $\mathcal{U}_1$'s last instruction. The other steps are detailed below:

\subsubsection{Locating the shellcode and jump over the encoded payload}
\label{sec:jal}
To make the shellcode position independent, we find its absolute position in memory using the jump and link (\texttt{jal}) instruction which stores the program counter to a user-specified register. This instruction consequently increases the shellcode's size by jumping over a large memory region. Yet, this area is not entirely wasted, as we repurpose it to store our packed payload $\mathcal{P}$.

\subsubsection{Fixing-up the store pointer}
\label{sec:fixupstore}
The next step consists in setting up the register $\texttt{XI}$ containing the address at which we will write stage 2. For this purpose we use the absolute address obtained in \Cref{sec:jal}, to which we add a constant using several \texttt{addi} instructions. We must not forget the additional offset required when using the \texttt{sd} store instruction in the decoder. Consequently, stage 2 will be unpacked immediately after the shellcode.

The biggest immediate available for the \texttt{addi} instruction in $\alpha\texttt{RV64GC}$ is \texttt{464}. Since the shellcode is much longer, we use the following trick: we first append several \texttt{addi XI, XI, 464} instructions until we exceed the desired value. Then we replace some immediates in the sequence by the second greatest available immediate, i.e. \texttt{448}, which reduces the total sum, until the desired value is reached.\footnote{A small \texttt{NOP} sled of at most 16 bytes may be required for getting an exact match.} In this way, we are guaranteed to use the least amount of \texttt{addi} instructions possible.

\subsubsection{Unpacking stage 2}
\label{sec:stage2unpack} \label{sec:loadgen}
We then unpack stage 2 starting at $\texttt{XI} + \textit{store\_offset}$, where $\texttt{XI}$ is the register we set previously. This is done sequentially, using the \texttt{sd} instruction with carefully chosen offsets. Indeed, we have many offsets only 2 bytes apart. In our case, we chose a long chain of offsets (available from our constrained instruction set), each exactly 2 bytes apart, \texttt{1920}, \texttt{1922}, ..., \texttt{1938}. This allows storing at most 20 consecutive bytes by first loading 2 bytes into a register and then storing them into memory. We use a precomputed table providing for each immediate the minimal sequence of instructions needed for loading it to a given register. We explain below how to compute this table. To store more than 20 bytes, we increment \texttt{XI} (using the \texttt{addi XI, XI, 16} instruction) between each batch of 16 bytes, and continue with offsets \texttt{1924}, ..., \texttt{1938}. The whole stage 2 is 40 bytes long, unpacked in 3 batches of 20, 16, and 4 bytes.

The above strategy relies on a precomputed table of sequences for achieving arbitrary 2-byte loads. We generate this table using a depth-first search strategy, by iterating over \alphash instruction sequences and storing the reached values. This approach yields for each 2-byte immediate the shortest sequence required to load it into a register.

More precisely, the first instruction of the sequence is a \texttt{lui} (loads an immediate in bits 12 to 31 of the destination register). It is followed by an arithmetical right-shift \texttt{sra} instruction (unless the shift amount is null). By intersecting the set of possible registers which may be used both as destination register for \texttt{sd} and \texttt{lui}, we end up with registers \texttt{s4}, \texttt{s6}, \texttt{t1}, \texttt{tp}. As \texttt{sra} requires a register as a shift amount, we also iterate over all possible load immediate \texttt{li} and \texttt{addiw} subsequences to get the desired shift amount.

The next instructions of the sequence are made of \texttt{addiw} instructions, with immediates ranging from \texttt{-20} to \texttt{30}. We limit the exploration of the instruction sequence space to at most 4 \texttt{addiw} instructions, to keep $\mathcal{U}_1$ compact. This limitation still grants the possibility to load 63448 out of the 65536 possible values (or 96\,\%) into \texttt{s4}, \texttt{s6}, \texttt{t1}, or \texttt{tp}.

In this way, we can design our stage 2 with a substantially expanded set of available instructions. Indeed, we merely require to make sure every pair of bytes in stage 2 could be loaded from an instruction sequence in the table.

\subsection{Stage 2}
\label{sec:stage2}
Stage 2 ($\mathcal{U}_2$) is more straightforward. It consists of some initialization code followed by a loop whose body decodes two consecutive bytes of $\mathcal{P}_{enc}$, the encoded payload. The full implementation can be found in \Cref{appendix:sourcecode}. The initialization code sets three registers --- the \emph{reading pointer} \texttt{XP} pointing to the encoded payload, the \emph{writing pointer} \texttt{XQ} pointing to the start of the decoded payload, and the \emph{end pointer} \texttt{XS} pointing to the end of the decoded payload. For simplicity, $\mathcal{U}_2$ performs in-place decoding, meaning that \texttt{XP} is initially equal to \texttt{XQ}.

We also flush the instruction cache with a \texttt{fence.i} instruction, which is required as we modify executable memory. We discuss later in \Cref{sec:eval} the assumption that the first \texttt{fence.i} is not shadowed in the instruction cache.

Since 63 characters are available, it is theoretically possible to encode almost 6 bits of the payload in a single alphanumeric byte of the shellcode. However, to keep $\mathcal{U}_2$ short, we decided to encode only 4 bits per alphanumeric byte. This spreads each byte of the payload over 2 consecutive alphanumeric characters. As stage 2 is unpacked sequentially by the first stage, we need to make stage 2 the shortest possible, even if this makes the encoder more complex. Indeed, any additional length here would lead to a significant increase in stage 1 size. Let $K$ be the byte stored at $\texttt{XP}+1$, $L$ the byte stored at \texttt{XP} and $A$ the byte written at address \texttt{XQ} by the store instruction. The decoding algorithm we devised only requires 5 instructions in the body of its loop.

\newcommand{\colorListingsComment}{DarkGreen}
\begin{lstlisting}
lw   XS, 4(XP)  #(*@\textcolor{\colorListingsComment}{{ Load $K$ and $L$ bytes}}@*)
#(*@\textcolor{\colorListingsComment}{{ XS == 0x????$K_{[4:7]}K_{[0:3]}L_{[4:7]}L_{[0:3]}$}}@*)
mv   XT, XS     #(*@\textcolor{\colorListingsComment}{{ Duplicate value}}@*)
srli XT, XT, 4  #(*@\textcolor{\colorListingsComment}{{ Shift right by 4}}@*)
#(*@\textcolor{\colorListingsComment}{{ XT == 0x?????$K_{[4:7]}K_{[0:3]}L_{[4:7]}$}}@*)
xor  XS, XS, XT #(*@\textcolor{\colorListingsComment}{{ XS := XS $\bm{\oplus}$ XT}}@*)
#(*@\textcolor{\colorListingsComment}{{ XS == 0x??????$A_{[4:7]}A_{[0:3]}$}}@*)
sw   XS, 0(XQ)  #(*@\textcolor{\colorListingsComment}{{ Store decoded byte $A$}}@*)
\end{lstlisting}

Hereafter, we find the encoding formulae by solving the decoding equations. Henceforth, when encoding byte $A$, the encoder must find values for $K$ and $L$ so that:
\begin{align*}
&K~\textnormal{and}~L~\textnormal{are}~\textnormal{alphanumeric}\\
&L\;[0:3] \bm{\oplus} L[4:7] = A[0:3]\\
&K[0:3] \bm{\oplus} L[4:7] = A[4:7]
\end{align*}
One should remark that every byte of the form \texttt{0x4*} or \texttt{0x6*} for \texttt{*} non null is alphanumeric. This simplifies the resolution of the previously given constraints. The following solution can be checked to give an alphanumeric encoding for any input byte.
\begin{align*}
L[4:7] &= \texttt{0x4} ~\textnormal{if} ~ A[0:3] \neq \texttt{0x4} ~\textnormal{else} ~\texttt{0x6}\\
L[0:3] &= A[0:3] \bm{\oplus} L[4:7]\\
K[0:3] &= A[4:7]\bm{\oplus} L[4:7]\\
K[4:7] &= \texttt{0x4} ~\textnormal{if} ~ A[0:3] \neq \texttt{0x0} ~\textnormal{else} ~\texttt{0x5}
\end{align*}

Finally, as executable memory modifications occurred, we flush the instruction cache again using a \texttt{fence.i} instruction, and jump to the decoded payload $\mathcal{P}$.

\subsection{Payload}
\label{sec:payload}
Stage 3 is the payload $\mathcal{P}$, containing arbitrary binary code. We generate this code directly from a C source payload compiled with a standard gcc. The resulting binary code is then encoded as described in \Cref{sec:stage2} with a lightweight PHP script.

The size of the payload is upper bounded by the offset chosen for the forward jump is \Cref{fig:stage1}. For our needs, we deemed 1024 bytes to be sufficient, allowing us a decoded payload of 512 bytes. Note that with some minor engineering work, this maximum size can be increased. In the context of usual shellcoding attacks, the payload almost always fits into this limit. As a proof-of-concept, we test in \Cref{sec:eval} three different payloads for a standard Linux: a \texttt{printf("Hello world")} shellcode, an \texttt{execve("/bin/sh")} shellcode, and one that leaks the contents of \texttt{/etc/shadow}.

\subsection{Integration/Linking}
All in all, the complete shellcode is built in the following order:
\begin{enumerate}
	\item We compute the table of minimal instruction sequences (\Cref{sec:loadgen}).
	\item We build the final payload $\mathcal{P}$, and compute its length (\Cref{sec:payload}).
	\item We generate stage 2, with the appropriate values for the reading pointer, the writing pointer, and the end pointer (\Cref{sec:stage2,fig:stage2}).
	\item We generate the unpacker for stage 2, and compute its length (\Cref{sec:stage2unpack,fig:stage1}).
	\item We generate the code for fixing-up the store pointer (\Cref{sec:fixupstore}).
	\item We then build the whole shellcode, without its encoded stage 3 payload, for which we allocated the necessary space.
	\item We finally insert the encoded payload $\mathcal{P}$ at the appropriate location in the shellcode.
\end{enumerate}

\subsection{Shellcoding in \alpslash}
\label{sec:alpslash}
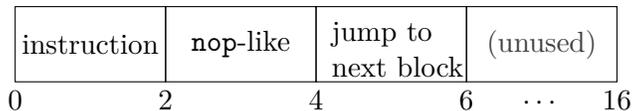
\begin{figure}[ht]
	\centering
	\begin{tikzpicture}
	\draw (0,0) rectangle (2, -1);
	\draw (2,0) rectangle (4, -1);
	\draw (4,0) rectangle (6, -1);
	\draw (6,0) rectangle (8, -1);
	
	\node[] at (1,-.5) {instruction};
	\node[] at (3,-.5) {\texttt{nop}-like};
	\node[below right,text width=1.85cm] at (4.075,-0.075) {jump to next block};
	\node[darkgray] at (7,-.5) {(unused)};
	
	\node[above] at (0,-1.5) {0};
	\node[above] at (2,-1.5) {2};
	\node[above] at (4,-1.5) {4};
	\node[above] at (6,-1.5) {6};
	\node[above] at (7,-1.5) {$\cdots$};
	\node[above] at (8,-1.5) {16}; 
	\end{tikzpicture}
	\caption{Diagram of a 16-byte block. Our stage 2 instructions are located in the first two bytes, while the next two contain a \texttt{NOP} like instruction followed by a jump to the next block. The last 10 bytes are unused.}
	\label{fig:block}
\end{figure}

We have also created a version of the shellcode in \alpslash, using atomic store instructions instead of regular stores for unpacking in stage 1. Data is stored with the \texttt{amoor.d} instruction which operates on 8 naturally aligned bytes. By opposition to the previous implementation in \alphash, we do not have offsets for stores, hence we need to modify the store pointer using available \texttt{addi} instances, which can only increase a register by a multiple of 16.
We thus store our decoded stage 2 in blocks of 16 bytes.
As we have control over only the 8 first bytes, we decided to split them into two parts, the first four bytes containing the decoded instruction, whereas the next two bytes contain a jump instruction to the next block (\texttt{j .+0xc} , or \texttt{0x31A0} in hexadecimal). The structure of the block is shown in Figure~\ref{fig:block}.

As required by the sequences for 2-byte arbitrary load computed in \Cref{sec:loadgen}, we wrote stage 2 using only compressed instructions.
The only exception is \texttt{fence.i}, which is unavoidable and does not have a compressed version.
In this case, we use a custom sequence to store its value (\texttt{0x0000100F} in hexadecimal).
We would like to particularly thank the authors of RISC-V for the fact that the 16 highest bits of \texttt{fence.i} are all zeros, which keeps our sequence of instructions really short.
Otherwise we would have required chaining many \texttt{addi} instructions, making the shellcode too long to be used in practice.

The sequences used for loading the 2-byte instructions are computed using a table similar to that of \Cref{sec:loadgen}.
By opposition to \alphash, here the word's two highest bytes will be executed as an instruction. We make sure that these two bytes do not modify the high-level semantics of the program.
Altogether, the table allows loading 58174 possible 16-bit values, out of 65536 (or 88\,\%) which still allows encoding our stage 2 with only minor modifications, at the expense of a slight size increase of only 2 bytes. The payload and the way it is encoded remains identical.

\subsection{Shellcoding in \alptick}
Shellcoding in \alptick is more tricky. First of all, it requires the \emph{floating-point unit} (FPU) to be activated, which is always done by the operating system when working in a hosted environment.
In the context of the bare-metal examples presented in this paper, we use a small additional piece of non-alphanumeric code, whose sole purpose consists in activating the FPU (in big-endian hexadecimal: \texttt{0x896373900330}).

Similarly to \alpslash, the main difference lies in the way stage 2 is unpacked by $\mathcal{U}_1$. This time, we store in the data-pool some floating-point values which are used by $\mathcal{U}_1$ during unpacking. The most general floating-point data manipulation instruction available is \texttt{fmadd r, a, b, c} (fused-multiply add): it computes $r = a\times b+c$. The store operation \texttt{fsd} then stores $r$ at the desired memory location. We thus have to solve equations of the form $r_i = a_i \times b_i + c_i$, where $r_i$ is a small part of the decoded stage 2, under the constraint that each $a_i$, $b_i$ and $c_i$ need to be loaded from the data pool. To keep our data pool as small as possible, we need to share values between different equations. As this increases the mathematical complexity of solving floating-point equations, we decided to work on a simplified version of the problem, in which we only encode 6 bytes of stage 2 into $r_i$. Indeed, in this way, the constraint lies only in the mantissa of the floating-point. Furthermore, we fixed the two remaining bytes of $a_i$, $b_i$ and $c_i$ to alphanumeric constants which do not propagate carries to the exponent when performing the operation. These simplifications turned out to be sufficient for solving the equations while reducing the total number of different floating-point constants.

Hereafter, we present some of the methods we used for reducing the number of different floating-point constants used. The first, consists in using the same $b_i$ for all equations, as, without loss of generality, this does not impedes finding a solution by just modifying $a_i$ and $c_i$. This simplification allows solving the equation by testing random alphanumeric values for $a_i$, computing the adequate $c_i$ then checking both $a_i$ and $c_i$ are alphanumeric. A simple combinatorial analysis gives us the approximate probability that a randomly chosen alphanumeric $a_i$ gives an alphanumeric solution for $c_i$ as: $(\frac{62}{256})^6 \simeq \frac{1}{50000}$.

The second method consists in using the same $a_k$ for two consecutive equations. Formally, we require to find solutions $a_{k}$, $c_{2k}$, $c_{{2k}+1}$ for the following set of equations:
\begin{align*}
r_{2k} &= a_k \times b + c_{2k} \\
r_{2k+1} &= a_k \times b + c_{2k+1}
\end{align*}

Unfortunately, these equations are not guaranteed to always have solutions. Indeed, let $r_{2k}$ and $r_{2k+1}$ differ in their highest bit. This means the highest bits of $c_{2k}$ and $c_{2k+1}$ are different.\footnote{We omit the rare and lucky case where carry propagation still provides a solution to the equation.} Hence one of them is non-alphanumeric. The solution we came up with consists in making stage 2 polymorphic, and trying to solve those equations for all instances of stage 2, hoping to find one for which all the equations have a solution. The different stage 2 instances are generated by either modifying the registers (150k variants), reordering some initialization instructions for the loop (6 variants), or reordering the pointer increment instruction in the loop's body (7 variants); yielding a total of about 6 millions stage 2 instances.

\begin{algorithm}[!b]
	\KwIn{$b$, a 64-bit floating-point value}
	\KwIn{$s_0, ..., s_{2\ell+1}$, the stage 2}
	\KwResult{a list of 64-bit floating-point values}
	mem := Array(\texttt{None}) \;
	P := Polymorphism($s_0, ..., s_{2\ell+1}$) \;
	\ForEach{$r_0, ..., r_{2\ell+1}$ \textnormal{\textbf{in}} $P$}
	{
		\For{$k=0~\textnormal{\textbf{to}}~\ell$}
		{
			\If{$mem[r_{2k}][r_{2k+1}]$ \textnormal{\textbf{is not}} \textnormal{\texttt{None}}}
			{
				\textnormal{\textbf{continue}}
			}
			\For{$i = 0~\textnormal{\textbf{to}}~2000000$}
			{
				$a$ := RandAlphanumFloatingPoint()\\
				Solve $c_{2k}$ in \\ \qquad$r_{2k} = a \times b + c_{2k}$\\
				Solve $c_{2k+1}$ in \\ \qquad $r_{2k+1} = a \times b + c_{2k+1}$\\
				\If{$c_{2k}$ and $c_{2k+1}$ \textnormal{are alphanumeric}}
				{
					mem[$r_{2k}$][$r_{2k+1}$] := $a$\\
					\textnormal{\textbf{break}}
				}
			}
			\If{$mem[r_{2k}][r_{2k+1}]$ \textnormal{\textbf{is}} \textnormal{\texttt{None}}}
			{
				$mem[r_{2k}][r_{2k+1}]$ := $\texttt{NotFound}$
			}
		}
		\If{$\not\exists k, mem[r_{2k}][r_{2k+1}] \textnormal{\textbf{ is }} \textnormal{\texttt{NotFound}}$}
		{
			\textnormal{\textbf{return}} $(mem[r_{2k}][r_{2k+1}])_{(k=0..\ell)}$
		}
	}
	\caption{Automated testing of the existence of a solution to the sets of equations induced by a specific stage 2 encoding. The outer loop is parallelized, testing several stage 2 instances concurrently.}
	\label{alg1}
\end{algorithm}

Algorithm~\ref{alg1} uses memoization to speed up the resolution of equations. In the worst case, the first loop has 12 million iterations (which can be executed in parallel), the second has 4 iterations while the last has 2 million iterations. In practice, when taking into account memoization, we counted $2.3 \times 10^{11}$ iterations, requiring 1.5 execution hours on a 4-core Atom 2\,GHz CPU. Eventually, we found several instances for which all equations had a solution. The rest of the shellcode is built in the same fashion as the previous versions presented in the previous sections.

\section{Evaluation}
\label{sec:eval}
\subsection{QEMU}
We initially tested our 3 shellcodes on QEMU~\cite{QEMU}, a widespread open-source emulator. It emulates a HiFive Unleashed \texttt{RV64GC} development board, without some of its micro-architectural features like caches or timings. The payload is expected to print ``\textit{Hello world!}'' on the serial device mapped at address \texttt{0x10013000}. After generating the corresponding shellcodes for \alphash, \alpslash and \alptick, we successfully managed to execute them on QEMU. We provide in \Cref{appendix:helloworld} the generated shellcodes, as well as some instructions to easily reproduce this experiment.

\subsection{HiFive Unleashed}
Subsequently, we moved to a more realistic environment, including a Linux operating system on a HiFive Unleashed board powered by a quad-core Freedom U540 \texttt{RV64GC} processor. It features an off-the-shelf Fedora 28 stage 4 disk image in a buildroot chrooted environment, for which we created a purposely vulnerable application executing its input data.

The first payload uses the \texttt{write} system call to print ``\textit{Hello world!}'' on the standard output. As previously, we generate the three different versions of our shellcode, and successfully manage to execute them on the vulnerable application. We successfully test the three shellcodes with two other payloads, one that spawns a shell using the \texttt{execve} system call, and one that prints on the standard output the contents of \texttt{/etc/shadow} file, using the \texttt{openat}, \texttt{read} and \texttt{write} system calls.

As a side note, as the floating-point unit is activated by the operating system, our \alptick shellcode does not require the non-alphanumeric previously described gadget anymore. Furthermore, we did not observe any instruction cache issue, as one could dread when using self-modifying code. This can be explained by the use of \texttt{fence.i} instructions that synchronize the instruction cache.

\section{Conclusion and future work}
We described a methodology for writing arbitrary alphanumeric (+1) RISC-V shellcodes. This method relies on unpacking, in which a program written in a very constrained instruction set stores into the memory another program written in a less constrained instruction set. Here, we required two unpackers in a three-staged shellcode to achieve arbitrary code execution. As a proof-of-concept, we showed examples of such shellcodes for the HiFive Unleashed board, featuring a standard Linux operating system. These positive results validate our choice for unpacking methods as the most suitable solution to the problem of writing executable code in a very constrained ISA subset.

Besides, the shellcodes provided in this paper only show proof-of-concept attacks. With the wide adoption of RISC-V based devices, we expect the attack surface to widen as new applications are published. Thereupon, we hope to see adequate defense mechanisms being deployed on those platforms, preventing such an attack. On the attacking side, automation seems the most promising way of improvement. Indeed, shellcodes tend to be handwritten or automated using \textit{ad hoc} algorithms. We believe that a more general approach based on a higher-level semantic representation of the available instructions may be able to comprehensively solve the problem of writing code in a constrained ISA subset.
\printbibliography[]

\pagebreak

\appendix

\section{Hello World Shellcodes}
\label{appendix:helloworld}

\newcommand{\colorUnassigned}{black}
\newcommand{\colorNameUnassigned}{black}
\newcommand{\colorInitJump}{FireBrick}
\newcommand{\colorNameInitJump}{red}
\newcommand{\colorLoadPool}{Green}
\newcommand{\colorNameLoadPool}{green}
\newcommand{\colorPayloadEnc}{DarkBlue}
\newcommand{\colorNamePayloadEnc}{blue}
\newcommand{\colorMoveLinkToSp}{\colorInitJump}
\newcommand{\colorNameMoveLinkToSp}{\colorNameInitJump}
\newcommand{\colorStageTwoMajic}{DarkOrange}
\newcommand{\colorNameStageTwoMajic}{orange}
\newcommand{\colorSpFixup}{cyan}
\newcommand{\colorNameSpFixup}{cyan}
\newcommand{\colorDecodeStageTwo}{violet}
\newcommand{\colorNameDecodeStageTwo}{purple}
\newcommand{\colorEndNopsled}{brown}
\newcommand{\colorNameEndNopsled}{brown}

We provide ready-to-use demo shellcodes, written respectively in \alphash, \alpslash and \alptick.
They print ``\textit{Hello world!}'' on the serial output, when executed on QEMU with the following command:

\noindent\texttt{ qemu-system-riscv64 -nographic -machine sifive\_u\hfill\null} \\
\texttt{\null\hfill{}-device loader,file=shellcode.bin,addr=0x80000000}

The notation \texttt{(X)\^{}\{Y\}} means that \texttt{X} is repeated \texttt{Y} times.

Colors have been added to each shellcode, with each color describing a specific high-level operation described in section \ref{sec:Implementation}.
The instructions that jump over the encoded payload and put the location of the shellcode in \texttt{sp} as described in section~\ref{sec:jal} are colored in \textcolor{\colorInitJump}{\colorNameInitJump}.
The encoded payload is in \textcolor{\colorPayloadEnc}{\colorNamePayloadEnc}.
Fixing-up the store pointer (section~\ref{sec:fixupstore}) is in \textcolor{\colorSpFixup}{\colorNameSpFixup}.
Unpacking the stage 2 (section~\ref{sec:stage2unpack}) is in \textcolor{\colorDecodeStageTwo}{\colorNameDecodeStageTwo}.
The final nopsled is in \textcolor{\colorEndNopsled}{\colorNameEndNopsled}.
For \alpslash and \alptick, additional data stored in the data pool is shown in \textcolor{\colorLoadPool}{\colorNameLoadPool}.
In \alpslash, some additional code is required to first store the jump instruction (as shown in section~\ref{sec:alpslash}) which is here in \textcolor{\colorStageTwoMajic}{\colorNameStageTwoMajic}.
Unused parts of the shellcode are in \textcolor{\colorUnassigned}{\colorNameUnassigned}. Given that colors cannot be reproduced in print, we refer the reader to the Arxiv version of this paper to that end.

\subsection{\alphash QEMU Hello World}
\begin{center}
	\texttt{ \ \\
		\textcolor{\colorInitJump}{o\#0\#}{\space(BBBB)\^{}\{1304\}\space\space}\textcolor{\colorPayloadEnc}{CGEDEDDDOEEDEEDDGEEE} \\
		\textcolor{\colorPayloadEnc}{ECEDGEEDEDLAKJDDDBDDEDDNCMCDDDDDGMCLCFFD} \\
		\textcolor{\colorPayloadEnc}{COBG}\textcolor{\colorPayloadEnc}{EDDEGDCHCDDDALCDLMFHGDCHCDDDACOKEDAP} \\
		\textcolor{\colorPayloadEnc}{FLDLDDDDDDDDLPABHBHBKBHFDFCC}\textcolor{\colorPayloadEnc}{KBFCHBbPEFND} \\
		\textcolor{\colorPayloadEnc}{DDDD}\textcolor{\colorUnassigned}{BB}{\space(BBBB)\^{}\{1377\}\space\space}\textcolor{\colorMoveLinkToSp}{3Z0A3QCA}\textcolor{\colorSpFixup}{yayayayaya} \\
		\textcolor{\colorSpFixup}{yayayayayayayayayayayayayayaya}\textcolor{\colorDecodeStageTwo}{EcY3e\#\#0ax} \\
		\textcolor{\colorDecodeStageTwo}{Aj\#1Ay75v71J3SEAi\#\#2ax7Eo91J3SEAY\#\#3ax75} \\
		\textcolor{\colorDecodeStageTwo}{\#zMJ3SEAM\#y\#y\#}\textcolor{\colorDecodeStageTwo}{\#4axQcY3E\#\#5ax7ER81J3SEAY\#} \\
		\textcolor{\colorDecodeStageTwo}{\#6ax7Ej81J3SEAY\#\#7ax75PP9J3ZEA\#8Ay7\#z8}\textcolor{\colorDecodeStageTwo}{1I} \\
		\textcolor{\colorDecodeStageTwo}{3Z\#A\#9AyAa75r05J3ZEA\#2Ay7EBA9J3ZEA\#3Ay7\#} \\
		\textcolor{\colorDecodeStageTwo}{F\#1Im93S\#Au3\#4ax7Ea85J}\textcolor{\colorDecodeStageTwo}{3SEAY3\#5ax7Up01J3Z} \\
		\textcolor{\colorDecodeStageTwo}{EA\#6Ay759M5J3SEAi\#\#7axAcy3e3\#8axEcY3e\#\#9} \\
		\textcolor{\colorDecodeStageTwo}{axAaAj}\textcolor{\colorDecodeStageTwo}{\#2Ay7\#h91I3Z\#A\#3Ay}\textcolor{\colorEndNopsled}{ySySySySs0A4}\phantom{aaaa}
	}
\end{center}
\pagebreak
\subsection{\alpslash QEMU Hello World}
\begin{center}
	\texttt{ \ \\
		\textcolor{\colorInitJump}{ySySo/0/}\textcolor{\colorLoadPool}{BBBBB03JBBBBBBBBBBBBBPCJ}\phantom{aaaaaaaa} \\
		{\space(BBBB)\^{}\{1955\}\space\space}\phantom{aaaaaaaaaaaaaaaaaaaaaaaaa} \\
		\textcolor{\colorPayloadEnc}{CGEDEDDDOEEDEEDDGEEEECEDGEEDEDLA}\textcolor{\colorPayloadEnc}{KJDDDBDD} \\
		\textcolor{\colorPayloadEnc}{EDDNCMCDDDDDGMCLCFFDCOBG}\textcolor{\colorPayloadEnc}{EDDEGDCHCDDDALCD} \\
		\textcolor{\colorPayloadEnc}{LMFHGDCHCDDDACOK}\textcolor{\colorPayloadEnc}{EDAPFLDLDDDDDDDDLPABHBHB} \\
		\textcolor{\colorPayloadEnc}{KBHFDFCC}\textcolor{\colorPayloadEnc}{KBFCHBbPEFNDDDDD}{\space(BBBB)\^{}\{751\}}\phantom{aaa} \\
		\textcolor{\colorMoveLinkToSp}{3Y0A3Q/A}\textcolor{\colorStageTwoMajic}{Bj/8Aa/8Aa1J3RHA3Z0A/0Ac/8AD//Aa} \\
		\textcolor{\colorStageTwoMajic}{/2AA}\textcolor{\colorSpFixup}{9a9a9a9a9a9a9a9a9a9a9a9a9a9a9a9a9a9a} \\
		\textcolor{\colorSpFixup}{9a9a9a9a9a9a9a9a}\textcolor{\colorDecodeStageTwo}{3Z0A/0Ac/8AD75/AIJ3SEA13} \\
		\textcolor{\colorDecodeStageTwo}{1313//aDAa3Z0A/0Ac/8AD75xG1J3SEAi3//aDAa} \\
		\textcolor{\colorDecodeStageTwo}{3Z0A/0Ac/8AD7EqI1J3SEAY3//aDAa3Z0A/0Ac/8} \\
		\textcolor{\colorDecodeStageTwo}{AD7EpQ9J3ZEA//AAAa3Z0A/0}\textcolor{\colorDecodeStageTwo}{Ac/8AD75gA9J3SEA} \\
		\textcolor{\colorDecodeStageTwo}{y3//aDAa3Z0A/0Ac/8AD7ETH1J3SEAY3//aDAa3Z} \\
		\textcolor{\colorDecodeStageTwo}{0A/0Ac/8}\textcolor{\colorDecodeStageTwo}{AD7ElH1J3SEAY3//aDAa3Z0A/0Ac/8AD} \\
		\textcolor{\colorDecodeStageTwo}{75PP9J3ZEA//AAAa3Z0A/0Ac/8AD7/zH}\textcolor{\colorDecodeStageTwo}{1I3Z/A//} \\
		\textcolor{\colorDecodeStageTwo}{AAAa3Z0A/0Ac/8AD75r05J3ZEA//AAAa3Z0A/0Ac} \\
		\textcolor{\colorDecodeStageTwo}{/8AD7EBA9J3ZEA//}\textcolor{\colorDecodeStageTwo}{AAAa3Z0A/0Ac/8AD7/F/1Im9} \\
		\textcolor{\colorDecodeStageTwo}{3S/Au3//aDAa3Z0A/0Ac/8AD7Ea85J3SEAY3//aD} \\
		\textcolor{\colorDecodeStageTwo}{Aa3Z0A/0Ac/8AD7UpP1J3ZEA//AAAa3Z0A/0Ac/8} \\
		\textcolor{\colorDecodeStageTwo}{AD75aA1J3SEAY3A3//aDAa3Z}\textcolor{\colorDecodeStageTwo}{0A/0Ac/8AD7///1I} \\
		\textcolor{\colorDecodeStageTwo}{a93S/AY3M31313//aDAa3Z0A/0Ac/8AD75/AIJ3S} \\
		\textcolor{\colorDecodeStageTwo}{EA131313}\textcolor{\colorDecodeStageTwo}{//aDAa3Z0A/0Ac/8AD7/h91I3Z/A//AA} \\
		\textcolor{\colorDecodeStageTwo}{Aa}\textcolor{\colorEndNopsled}{ySySySySySySySs0A4}\phantom{aaaaaaaaaaaaaaaaaaaa}
	}
\end{center}

\subsection{\alptick QEMU Hello World}
\begin{center}
	\texttt{ \ \\
		{\textbackslash{}89\textbackslash{}63\textbackslash{}73\textbackslash{}90\textbackslash{}03\textbackslash{}30}\phantom{aaaaaaaaaaaaaaaaaaaaaa} \\
		\textcolor{\colorInitJump}{o'0'}\textcolor{\colorUnassigned}{BBBBBBBBBBBBBBBBBBBBBBBBBBBBBBBBBBBB} \\
		\textcolor{\colorUnassigned}{BBBBBBBBBBBB}\textcolor{\colorLoadPool}{3B1ozDaBBZzqspbB}\textcolor{\colorUnassigned}{BBBBBBBBBBBB} \\
		\textcolor{\colorUnassigned}{BBBB}\textcolor{\colorLoadPool}{64cinpaB}\textcolor{\colorUnassigned}{BBBBBBBBBBBBBBBB}\textcolor{\colorLoadPool}{ug51zDaBVIQn} \\
		\textcolor{\colorLoadPool}{4f1A1nKj52aB}\textcolor{\colorUnassigned}{BBBBBBBBBBBBBBBBBBBBBBBBBBBB} \\
		\textcolor{\colorUnassigned}{BBBB}\textcolor{\colorLoadPool}{phYdop1A9RlYo3aBPtIx'51AMKqGzV1A}\textcolor{\colorUnassigned}{BBBB} \\
		\textcolor{\colorUnassigned}{BBBBBBBBBBBBBBBBBBBBBBBBBBBBBBBBBBBBBBBB} \\
		\textcolor{\colorUnassigned}{BBBB}\textcolor{\colorLoadPool}{UUUUUU1ALR5eFXcB}{\space(BBBB)\^{}\{1177\}\space\space}\phantom{aaaa} \\
		\textcolor{\colorPayloadEnc}{CGEDEDDDOEEDEEDDGEEEECEDGEEDEDLAKJDDDBDD} \\
		\textcolor{\colorPayloadEnc}{EDDNCMCDDDDDGMCLCFFDCOBG}\textcolor{\colorPayloadEnc}{EDDEGDCHCDDDALCD} \\
		\textcolor{\colorPayloadEnc}{LMFHGDCHCDDDACOKEDAPFLDLDDDDDDDDLPABHBHB} \\
		\textcolor{\colorPayloadEnc}{KBHFDFCC}\textcolor{\colorPayloadEnc}{KBFCHBbPEFNDDDDD}\textcolor{\colorUnassigned}{BB}{}\phantom{\,\textbackslash}{(BBBB)\^{}\{1438\}}\\
		\textcolor{\colorMoveLinkToSp}{3Z0A3QGA}\textcolor{\colorDecodeStageTwo}{B5b6F'F8f4J9j1N2n3}\textcolor{\colorSpFixup}{yayayayaya9a9a} \\
		\textcolor{\colorSpFixup}{9a9a9a9a9a9a9a9a9a9a9a9a9a}\textcolor{\colorDecodeStageTwo}{C3A2'0azC3Ab'3} \\
		\textcolor{\colorDecodeStageTwo}{azG3Hr'6azG3HB'9azAa07X3L7G3IR'4azG3Ib'7} \\
		\textcolor{\colorDecodeStageTwo}{azG3GZ'9az}\textcolor{\colorEndNopsled}{s0A4}\phantom{aaaaaaaaaaaaaaaaaaaaaaaaaa}
	}
\end{center}

\section{Source code}
\label{appendix:sourcecode}
The full source code used for this paper is available at:\\
{\url{https://xn--fda.fr/riscv-alphanumeric-shellcoding}} \\
and \\
{\url{https://github.com/RischardV/riscv-alphanumeric-shellcoding}}.
\\ It contains all demos and tools used for this paper.
\end{document}